\documentclass[12pt]{article}
\usepackage{amssymb}
\usepackage{pslatex}
\usepackage{graphicx}
\usepackage{color}
\usepackage{subfigure}

\makeatletter

\@addtoreset{equation}{section}
\makeatother
\renewcommand{\arraystretch}{1.24}

\newcommand{\bea}{\begin{eqnarray}}
\newcommand{\eea}{\end{eqnarray}}
\newcommand{\be}{\begin{eqnarray}}
\newcommand{\ee}{\end{eqnarray}}
\newcommand{\nn}{\nonumber}

\def\tr{\mathop{\rm Tr}}

\begin{document}

\begin{titlepage}
\vskip1cm
\begin{flushright}
$\mathbb{UOSTP}$ {\tt 101101}
\end{flushright}
\vskip1.25cm
\centerline{
\bf
Symmetry Breaking Phase Transitions in ABJM Theory
with a Finite U(1) Chemical Potential}
\vskip1.25cm
\centerline{  Dongsu Bak$^a$, Kyung Kiu Kim$^b$ and  Sangheon Yun$^b$
}
\vspace{1.25cm}
\centerline{\sl a) Physics Department, University of Seoul, Seoul
130-743 {\rm KOREA}}
\vskip0.25cm
\centerline{\sl b) Institute for the Early Universe, Ewha Womans University, Seoul
120-750 {\rm KOREA}}
\vskip0.25cm
\centerline{\tt dsbak@uos.ac.kr, kimkyungkiu@gmail.com, sanhan@ewha.ac.kr }
\vspace{1.5cm}
\centerline{ABSTRACT}
\vspace{0.75cm}
\noindent
We consider the $U(1)$ charged sector of ABJM theory at finite temperature, which corresponds to the Reissner-Nordstrom AdS black hole in the dual type IIA supergravity description. Including back-reaction to the bulk geometry, we show that phase transitions occur to a broken phase where SU(4) R-symmetry of the field theory is broken spontaneously by the condensation of dimension one or two operators. We construct the composite operators out of fields in ABJM theory and describe the phase transition with the dual gravity solutions. We show numerically and analytically that the relevant critical exponents for the dimension one operator agree precisely with those of mean field theory in the strongly coupled regime of the large $N$ planar limit.

\vspace{1.75cm}
\end{titlepage}

\section{Introduction}
The ABJM theory is the three dimensional ${\cal N}=6\,\,$ $U(N)\times U(N)$ superconformal
Chern-Simons theory with level $(k, -k)$ and dual to the type IIA string theory on
AdS$_4\times \mathbb{CP}_3$ background \cite{ABJM}.
Some test of this duality has been carried out
largely based on the integrability with indication of some additional 
structures compared to the well known AdS$_5$/$CFT_4$ counterpart \cite{Minahan:2008hf}-\cite{Minahan2}.
The type IIA supergravity description is dual to the large $N$ planar limit
of the Chern-Simons theory  where one takes $N,\,\, k \rightarrow \infty$ while holding 
 't Hooft
coupling $\lambda= N/k$ fixed.

Some probes of the Chern-Simons plasma at finite temperature are
carried out 
 recently via the consistent,
$\mathbb{CP}_3$
invariant dimensional reduction of the type IIA supergravity \cite{BakYun}. Alternatively a
finite temperature plasma
can be completely characterized by towers of static length (1/mass) scales. They are arising as decaying
spatial length scales of a perturbation theory when local operators are inserted at a certain point of the
plasma. In Ref.~\cite{BakMin}, these scales are scanned for the low-lying bosonic modes from which the true mass gap $m_g$
(the lowest in all) and the Debye screening mass $m_D$ (the lowest in $CT$ odd sector of the theory \cite{Arnold:1995bh})
are found
for the Chern-Simons plasma. Including Yang-Mills plasmas, these two scales are well representing the universal
characteristics of a certain strongly coupled plasma. For instance, the ratio $m_D/m_g$ for the ${\cal N}=4$
SYM theory
and the two-flavor model of 
QCD are matching with each other in the strong coupling limit supporting such a picture \cite{Karch}.
Leading thermodynamic corrections of ABJM theory in the small $\lambda$ expansion are explored
in Ref.~\cite{Smedback:2010ji}.

In this paper, we note that the ABJM theory possesses $SU(4) \times U(1)$ R-symmetries
and consider its particular sector in which one turns on a finite $U(1)$ number
density. For the type IIA supergravity side,
this sector is described by the $U(1)$ charged
Reissner-Nordstrom (RN) AdS black brane solution whose physics
will be our basic concern for the study of the field theory  at a strong coupling \cite{BakYun}. We begin with
the thermodynamic stability of the RN AdS black brane solution and prove that
it is thermodynamically stable at all 
temperatures including the zero temperature
limit. This is contrasted with the R-charged black holes
in the type IIB supergravity which is dual to the R-charged sector of
the ${\cal N}=4$  super-Yang-Mills (SYM) theory. The R-charged solution becomes thermodynamically
unstable below a certain temperature and the validity of the solution itself will be
lost completely \cite{Son:2006em}. Thus the RN AdS solution of this note is, up to now,
the only known example where one has the thermodynamic
stability of the gravity solution at all  temperatures and the dual field theory description is precisely known
at the same time. In other words, we found that the RN AdS black brane solution is relevant dual gravity description for the finite temperature ABJM theory with a chemical potential in the whole temperature region including the zero temperature, which is not the case for the ${\cal N}=4$ SYM theory.

We go on to study phase structure and transitions occurring within this charged sector.
These phase transitions turn out to be a second order type where the $SU(4)$ R-symmetry is broken spontaneously \cite{Klebanov:1999tb}. For this  we scan behaviors of supergravity modes which are dual to the primary operators of some dimension $\Delta$ in the RN black hole background \cite{Nilsson:1984bj}. Among them, one finds
the bosonic modes with mass squared ranged over $-9/4 \le m^2 < -3/2$ which are responsible for the phase transition. As will be further explained later on, the only possibility for the present case is $m^2=-2$ corresponding to one $\Delta =1$ operator of $SU(4)$ representation {\bf 15} and two $\Delta=2$ operators of $SU(4)$ representations  {\bf 15} and {\bf 84}, whose detailed operator contents will be specified below. The transition related to the condensate of the $\Delta=1$ operator occurs at a critical temperature higher than that of the $\Delta =2$ operators. Above the transition temperature, the scalar field has to be set to zero to satisfy the required boundary condition while the other part of the original RN black brane solution remains intact. Below the transition temperature the scalar field begins to develop a profile whose boundary behavior represents a condensation of operator expectation value without introducing any external source field. Of course the geometry will then be back-reacted accordingly. The solution represents the condensation of the operator expectation values below the transition temperature, which takes a particular direction in the space of the $SU(4)$ representation {\bf 15}. Thus the $SU(4)$ symmetry is spontaneously broken down to its little group. 
We then study the critical exponents of the phase transitions  and show both analytically and numerically that the exponents precisely match with those of the mean field theory.
We shall also show that there is further symmetry breaking phase transition of the same nature
at a lower temperature which involves condensation of the $\Delta =2$ operators.

In Section 2, we shall discuss some relevant properties of the ABJM theory.
Section 3 deals with the gravity description of the RN black brane and related physics in the
field theory. 
In Section 4, our system described by  RN black brane
solution is shown to be thermodynamically stable.
In Section 5, we discuss the development of gravitational instabilities of scalar modes below the critical
temperature. This will fix the critical temperature which depends on the dimension of the corresponding
operators. In Section 6, we discuss the phase transition by studying the gravity solution representing
the condensation of operator expectation values including the back reaction to the bulk geometry. The critical
exponents are shown to agree precisely with those of the mean field theory. Last section is devoted to the interpretations
and concluding remarks.

\section{Thermodynamics of ABJM theory at small $\lambda$}
The on-shell degrees of the ABJM theory consist  of bosonic and fermionic matter fields
$Y^I$ and $\Psi_I\ (I=1,2,3,4)$ together with two gauge fields $A_m$ and $\bar{A}_m$.
The complex scalar fields $Y^I$ are in the representation of $({\bf N}, {\bf \bar{N}}, {\bf 4})$
of the $U(N) \times U(N)$ gauge as well as the $SU(4)$ R-symmetries. There are also the complexified Majorana fermions
$\Psi_I$, which are in the representation of $({\bf N}, {\bf \bar{N}}, {\bf \bar{4}})$. The
gauge fields $A_m$ and $\bar{A}_m$ in the adjoint representations of the first $U(N)$ and the
second $U(N)$ respectively, are coupled to the matter fields $\Phi=(Y^I\!,\,\, \Psi_I)$ by
\bea
D_m \Phi = \partial_m \Phi +i A _m\,\Phi -i \Phi\, \bar{A}_m\,.
\eea
For further details of the Lagrangian and notation used in this note, see for example Ref.~\cite{BakRey1}.
The theory possesses global ${\cal N}=6$ 3d superconfomal symmetries of $OSp(6|4)$ whose bosonic part is
given by the 3d conformal symmetry of $SO(3,2)$ multiplied by $SU(4)$ R-symmetry.  This corresponds to the isometry
of the AdS$_4 \times \mathbb{CP}_3$ in the type IIA supergravity side. But one crucial difference from the ${\cal N}=4$ case
is the fact that there is an extra global $U(1)$ symmetry. The relevant charge  is associated with the
$U(1)$ phase transformation of complex fields by a overall phase factor. Denoting overall $U(1)$ parts of gauge fields
by $A_m^{U(1)}$ and ${\bar{A}}^{U(1)}_m\!\!\!$,\,\,\, the above charge is only coupled to the
relative $U(1)$ gauge field $A_m^{U(1)}-{\bar{A}}^{U(1)}_m\!\!\!$.

Now note that the fields $Y^I$ and $\Psi_I$ carry the $U(1)$
charges $(-1,+1)$ or $-1$ in terms of the relative one. Due to the Gauss law constraint of the Chern-Simons theory,
these charges are always accompanied by $U(1)$ magnetic fluxes\footnote{The unit flux for the current case with $e=1$ is given by $h/e= 2\pi \hbar = 2\pi$ where we set $\hbar=1$.}
$ {2\pi\over k}\, (+1,-1)$.  Thus for general $k$ including even the nonabelian contributions, the basic degrees
in the field theory in the deconfined phase exhibit an anyonic nature due to the (generically nonabelian)
statistical interactions between them.
But due to the large $N$ planar limit 
where we send
$N$ and $k$ to infinity at the same time, the statistical interactions 
drop out since they are of higher order
in $1/k$. Therefore, for instance, the total effective number of degrees in the weakly coupled small $\lambda$ limit
is simply proportional to $N^2$. Namely the entropy density  at temperature $T$ takes a value \cite{ABJM}
\bea
{\cal S} (\lambda\rightarrow 0) ={21\zeta(3)\over \pi} N^2 T^2 
\eea
in the $\lambda \rightarrow 0$ limit. (The free energy density is always related to the entropy density by
${\cal F}= - {\cal S}\, T/3$ as dictated by the conformal symmetry.)
On the other hand in the strongly coupled large $\lambda$ region,
the system is described by the black brane solution in the gravity side whose Bekenstein-Hawking entropy density
reads \cite{ABJM}
\bea
{\cal S}(\lambda)= {16\pi^2\over 27 \sqrt{2}} {N^2 T^2\over \sqrt{\lambda}}\,.
\label{entropystrong}
\eea
The appearance of the ${1\over \sqrt{\lambda}}$ suppression factor is not understood from the direct computation
of the field theory. This drastic change of the number of degrees might be related to some remnant of
anyonic interactions but we do not have any supporting evidence for this picture.

Our main concern in this paper is the sector of the ABJM theory where one turns on a finite $U(1)$ number
density or equivalently the corresponding chemical potential $\mu$.
First consider the small $\lambda$ region of planar limit and the high enough temperature with
$T\gg \mu$. The fermions can be ignored in this limit. In the thermal circle compactified effective
theory, the fermions  can be integrated out at weak coupling and high
temperature limit as their Matsubara frequency starts
with $\pi T$. Of course the $U(1)$ current gets contributions from both bosonic and fermionic degrees
of the theory. Their currents are conserved not separately but only in sum. Hence at weak coupling of
small $\lambda$, instead of
building up a fermi surface, occupation of bosonic states is energetically preferred.
The scalar fields in this effective 2d description acquire a mass \cite{Yamada:2006rx},
\bea
m^2(T) = -\mu^2 + m_T^2\,,
\eea
where $m^2_T$ is the thermal mass correction. For $\lambda \ll 1$, the thermal
mass has the expression \cite{Smedback:2010ji}
\bea
m^2_T = {118\over 3}\, \lambda^2 (\log \lambda)^2\, T^2 + O(\lambda^2 \log\lambda)\,.
\eea
The theory lies in the unbroken phase if $m^2 (T) \ge 0$. On the other hand the system in the
symmetric phase becomes unstable when $m^2 (T)\ < \ 0$  or $ \mu\  \ll\  T\  < \ \sqrt{3\over 118}\,\, {\mu \over \lambda|\log\lambda|}$.
As argued in Ref.~\cite{Yamada:2006rx}, some of the operator of the field theory may acquire nonvanishing
expectation values possibly leading to $SU(4)$ R-symmetry broken phase. But the precise fate of the
system at weak coupling
requires a further study, which is beyond scope of the present paper.

\section{RN black brane and type IIA supergravity on AdS$_4 \times \mathbb{CP}_3$}
In the  strongly coupled region of large $\lambda$, the description
of the ABJM theory by the type IIA supergravity on AdS$_4\times \mathbb{CP}_3$ is appropriate
since geometry there is weakly curved. The type IIA  spetra compactified on $\mathbb{CP}_3$
space was known quite some time ago \cite{Nilsson:1984bj}. Each mode of the resulting 4d supergravity 
is dual to a gauge invariant primary operator whose scaling dimension $\Delta$ is protected against
quantum corrections. The presence of these bulk modes shows how the basic degrees of
freedom are organized in the strongly coupled side of the ABJM theory. Thus study of these  bulk modes
for a given supergravity background will be our main tool  probing the physics 
at the strong coupling.

The bulk modes consist of infinite towers of spectra from spin zero to spin two. We shall be here briefly
describing some of relevant low-lying modes for later discussions.  Let us begin with the case of spin
one: The lowest are two massless bulk gauge fields that are dual to the $\Delta =2$ current operators
in the field theory side. One is  for the current of $SU(4)$ singlet representation [\,$(000)$ in the $SU(4)$ Dynkin label
notation\,], which is identified with that of the extra $U(1)$ global symmetry. In the gravity side
the $U(1)$ is related to the $U(1)$ isometry of the M-theory circle from the 11d perspective and its charge is carried by
D0 branes in a rough sense. As was shown explicitly in Ref.~\cite{BakYun}, this gauge field is arising as
a linear combination of
\bea
A_\mu= A^{D0}_\mu + 3 A^{D4}_{\mu}\,,
\eea
where $A^{D0}_\mu $ and $A^{D4}_\mu $ respectively coupled to the D0 branes and $D4$ branes wrapping $\mathbb{CP}_2$
four cycle inside $\mathbb{CP}_3$. The other combination $\tilde{A}= A^{D0}_\mu - A^{D4}_{\mu}$ becomes massive
by Higgs mechanism with $m^2=12$, which couples to the $\Delta =5$ boundary current operator. It should be also noted
that the monopole operator with overall field theory  $U(1)$ charges $n(k,-k) \,\, (n\in {\bf Z})$ is the one example
of heavy BPS state coupled to this $U(1)$ bulk gauge field \cite{ABJM}.
The second massless gauge field is in the adjoint ${\bf 15}$  [$(101)$] of $SU(4)$  and couples to the
boundary $SU(4)$ current operator of dimension $\Delta =2$.

\begin{table}[ht]
{
\renewcommand{\arraystretch}{1.2}
\begin{tabular*}{153mm}{@{\extracolsep\fill}|l||l|l|l|l|l|}
\hline \hline
\phantom{aaaaaaaaaa}
& spin 0 & spin ${1\over 2}$
&  spin 1  & spin ${3\over 2}$ 
& spin 2  \\
\hline \hline
$\Delta=1$ & $(101)^+_{-2}$ &   &  & & \\
\hline
$\Delta={3\over 2}$ & & $(002)_{0}$ $(200)_{0}$ &   &  &  \\
$ $ & & $(010)_{0}$  &   &  &  \\
\hline
$\Delta=2$ &
$(202)^+_{-2}$ \ $(101)^-_{-2}$
&  & $(000)^-_{0}$ \ $(101)^-_{0}$ &   &\\
\hline
$\Delta={5\over 2}$ & & $(103)_{1}$ $(301)_{1}$ &   & $\,(010)_1\,$ &  \\
$ $ & & $(111)_{1}$  &   &  &  \\
\hline
  $\Delta=3$ & $(303)^+_{0}$ \ $(202)^-_{0}$   &
 & $(101)^-_{2}$ \ $(202)^-_{2}$
& &  $(000)^+_{0}$  \\
  &   $(400)^-_{0}$ $(004)^-_{0}$
& &
$(210)^-_{2}$  $(012)^-_{2}$
    & & \\
  &   $(210)^-_{0}$ $(012)^-_{0}$
&   &  & & \\
&   $(020)^-_{0}$
&   &  & & \\

\hline
{\small $\mathbb{CP}_3\,\,$} singlets
&  $(000)^+_{4}$ $(000)^-_{10}$ &
&  $(000)^-_{0}$ $(000)^-_{12}$ &
&  $(000)^+_{0}$      \\
&  $(000)^+_{18}$
&   &  & & \\
\hline
\end{tabular*}
\caption
    {\small The low lying spectra up to the operator
dimension 3 are presented. The upper and lower indices denote respectively
the parity and the mass squared value $m^2$ of the supergravity mode.
The whole spectra
of $\mathbb{CP}_3$
singlet sector are presented in addition.}
\label{tableone} }
\end{table}

For the bulk scalar modes, the lowest ones with $m^2=-2$ are relevant for our later discussions,
which correspond to $\Delta=1,\,\,2$ operators.  The $\Delta =1$ scalar mode couples to the
boundary operator of $SU(4)$ representation {\bf 15} [$(101)$] which takes the form
\bea
O^{I}_J = {\tr}\,\, Y^I Y^\dagger_J - (\rm trace\ part)\,,
\label{op1}
\eea
and has 15 independent components.
The $\Delta=2$ modes involve two field theory operators of $SU(4)$ representation
{\bf 15} [$(101)$] and {\bf 84} [$(202)$] whose operator contents read
\bea
&& \tilde{O}_{I}^J = {\tr}\,\, \Psi_I \Psi^{\dagger J} - (\rm trace\ part)\,, \nn\\
&& O^{IJ}_{KL} = {\tr}\,\, Y^{(I} Y^\dagger_{(K} Y^{J)} Y^\dagger_{L)}  - (\rm trace\ part)
\eea
where the trace part denotes any contractions between the upper and the lower indices.

The fermionic bulk modes start with $|m|=0$ corresponding to the operator dimension $\Delta={3\over 2}$:
These are boundary operators of $SU(4)$ representations {\bf 10} [$(002)$],   {\bf 6} [$(010)$],
and $\overline{\bf 10}$
[$(200)$]. The low lying modes up to $\Delta = 3$ are listed in Table \ref{tableone}.

A few comments are in order.
We note that there are no bulk supergravity modes that are charged under the massless $U(1)$ gauge
field. The knowledge about the compactification spectra tells us about only linearized fluctuations
of modes above the AdS$_4$ or the AdS$_4$ black brane solution. In Ref.~\cite{BakYun}, a consistent
$\mathbb{CP}_3$ compactification keeping all $SU(4)$ invariant modes is carried out explicitly.
Any solutions of this 4d system can be consistently embedded into the original 10d
supergravity theory.
%

Our starting Lagrangian for the further discussion is
\be
{\cal L}=\frac{1}{2 \kappa^2}  \Big(\mathcal{R} +6 - \frac{1}{4}F_{\mu\nu}F^{\mu\nu} -
\sum_{a=1}^{n}\big(\, (\nabla\phi_a)^2 + m_a^2\,\,\phi_a^2\,\big) 
\,\Big)\label{Lagrangian}
\ee
where\footnote{In the bulk, any dimensionful quantity is measured with respect to the AdS radius scale
$\ell$ which we set to be unity for the notational simplicity.}
\be
\kappa^{-2}= {N^2\over 6\pi \sqrt{2\lambda}}\,.
\label{newton}
\ee
In this action, the Einstein Maxwell part with the negative cosmological constant
is a fully consistent truncation \cite{BakYun}  while the remaining scalar part is only valid up to
quadratic order. Below we shall show that, above  the critical temperature of the $\Delta=1$ operator,
the stable solution of the system is given by the RN black brane solution with vanishing scalar fields.
This part of the solution is fully consistent as just stated.
Below the critical temperature, the relevant scalar
field begins to develop and the corresponding set of solutions is valid only if the magnitude of
the scalar field is small enough. But the set of solutions in the near critical region carries
all the information about universal natures of the phase transition including relevant critical exponents.

Similar models have been discussed many times in the bottom up approach
\cite{Horowitz einstein scalar,Horowitz criticality,Hartnoll:2008kx}. However, it is hard to identify  the
dual field theory and the corresponding operator for condensation in this bottom
up approach. In our work, on the contrary, the identification of the physics of boundary CFT is
straightforward.

Below we take the following ansatz,
\bea
ds^2&=&e^{2A(r)}\bigg(-h(r) dt^2 + dx^2 + dy^2\bigg)+\frac{dr^2}{h(r)}\,\,,\nonumber\\
A_{t}&=&A_{t}(r)\;,\;\;\;\;\phi_a=\phi_a(r)\,,
\label{metric1}
\eea
to describe a finite temperature system  with plane plus time ($\mathbb{R}^2\times \mathbb{R}$) translational
symmetries.
Plugging the above ansatz into the equations of motion, we are led to the following equations \cite{Gubser:2008pf}:
\bea
&& A''+\frac{1}{2}\sum_{i=1}^{n}\phi_i'^2=0\,\,,\nonumber\\
&& h''+3A'h'-e^{-2A}\;F_{tr}^2=0\,\,,\nonumber\\
&& (e^{A}\;F_{tr})'=0\;\,\,,\nonumber\\
&& h\phi_a''+(3A'h+h')\phi_a' - m_a^2\phi_a=0\,\label{equations of motion}
\eea
with a constraint,
\be
{h}\sum_{a=1}^{n}{\phi'}_a^2 -\frac{1}{2}e^{-2A}\;F_{tr}^2 -2A'h' -6h(A')^2 = -{6}+\sum_{a=1}^{n} m_a^2\phi_a^2\,\,.\label{constraint}
\ee
The third equation in (\ref{equations of motion}) can be solved by
\be
F_{rt} = 2q e^{-A(r)}\,,
\label{fund}
\ee
leading to the set of equations
\bea
&& A''+\frac{1}{2}\sum_{i=1}^{n}\phi_i'^2=0\,\,,\nonumber\\
&& h''+3A'h'-4 q^2e^{-4A}=0\,\,,\nonumber\\
&& h\, \phi_a''+(3A'h+h')\phi_a' - m_a^2\,\phi_a=0\,,\label{eqom}
\eea
which will be the starting point of our subsequent analysis.

The black brane solution,
\be
A(r)=r\,, \ \ h(r)= 1- e^{-3r+3r_H}\,,\ \ F_{rt}=\phi_a=0\,,
\ee
describes the uncharged sector of the ABJM theory 
at finite temperature.
Due to the conformal symmetry of the black brane background, the finite temperature phase here
depends on only one dimensionful  parameter which can be taken as the temperature $T$. Thus this
uncharged sector possesses only one finite temperature phase corresponding to the
high temperature limit. The temperature $T$ is identified with the Hawking temperature
of the black brane,
\be
T= {1\over 4\pi} h'(r_H) e^{A(r_H)}={3\over 4\pi} e^{r_H}\,,
\ee
where $r=r_H$ is the location of horizon. The Bekenstein-Hawking entropy density is given by
the expression ${\cal S}$ in (\ref{entropystrong}).
The energy density, the free energy density and the pressure are related to the entropy density
by ${\cal E}= {2\over 3} {\cal S}T= 2p = -2 {\cal F}$
as simply dictated by the  conformal symmetry of the background.
Some probes of system are investigated in Ref.~\cite{BakYun}  by studying
the response of the scalar and current operators to the external perturbation. The static length
scales including the true mass gap as well as the Debye mass are further studied in Ref.~\cite{BakMin}.

Our main interest of this note  is the RN black brane solution,
\bea
\label{ABJMcharged}
A(r)=r\;,\;\;\;\;h(r)=1-\epsilon e^{-3r}+q^2e^{-4r}\;,\;\;\;\;F_{rt}=2\,{q}\;e^{-r}\;,\;\;\;\;\phi_a=0\,,
\eea
which is an exact solution of the original 10d equations of  motion.
The parameters $\epsilon$ and $q$ here are respectively proportional to the mass  and the charge densities
of the RN black brane.
The horizon is located at $r=r_H$ with $h(r_H)=0$ satisfying explicitly
\be
\epsilon\, e^{-3 r_H}= 1+ q^2 e^{-4 r_H}\,.
\ee
The minimum of $h(r)$ occurs at
\be
e^{-r_m}={3\epsilon\over 4q^2}\,,
\ee
with $h'(r_m)=0$. No nakedness condition for the mass and charge  requires
$h(r_m) \ \le \ 0$ leading to the inequality
\be
\Big({\epsilon\over 4}\Big)^2  \ \ge \  \Big({q\over \sqrt{3}}\Big)^3\,,
\ee
where the inequality is saturated at zero temperature.
The Hawking temperature of the RN black brane becomes
\be
T={3\epsilon \, e^{-2 r_H}\over 4\pi}\Big(
1-{4q^2\over 3\epsilon}\, e^{-r_H}
\Big)\,.
\ee
The entropy, the energy and the charge densities read
\be
{\cal S}={2\pi\over \kappa^2}\, e^{2 r_H}\,,\ \ \ {\cal E}={\epsilon\over \kappa^2}\,,\ \ \
\rho={2q\over \kappa^2}\,,
\ee
with ${\cal F}={\cal E}-T{\cal S}$ and $\mu_G={\partial {\cal F}(T,\rho)\over \partial\rho}$
where $\mu_G$ is proportional to the field theory chemical potential $\mu$.
Hence the no nakedness condition now takes a form
\be
\Big({\kappa^2 \,{\cal E}\over 4}\Big)^2  \ \ge \  \Big({\kappa^2\,\rho\over 2\sqrt{3}}\Big)^3\,.
\ee
In Ref.~\cite{BakRey2},  this character is attributed to build-up of a fermi surface for a finite number of the
fermion number density. The weakly coupled massless fermions at finite temperature in general do satisfy
an analogous inequality with precisely the same powers. Further support of the  picture comes from the fact
that the specific heat at low enough temperature is linear as
\be
C_V= \gamma_V\,\, T\,,
\ee
which is another important characteristic  of the fermi surface.

However, there are some additional properties which cannot be understood from the
fermion picture. The entropy density even at zero temperature remains finite; namely,
\be
{\cal S}(T=0)= {\pi \over \sqrt{3}}\,\, \rho\,,
\ee
with a finite size of horizon at $e^{2r_H(T=0)}={q\over \sqrt{3}}$. Where this entropy comes from even at zero temperature
is not clear.  Recall further that the $U(1)$ current is conserved only in sum of bosonic and fermionic contributions together. Thus at weak coupling the build-up fermi surface would not be possible
if the bosons were
in an unbroken phase and, hence, putting charges to the bosonic states were
energetically favored. However  the symmetric unbroken phase
without condensation will be problematic as argued in the previous section  for the weakly coupled
case.
Therefore the system should be at some unbroken phase with some bosonic condensate at least for the
weakly coupled regime.
In later sections, we would like to study the nature of phases occurring at the RN black holes,
in the strongly coupled side, by the condensation of some operator expectation values.


Near boundary regions of large $r$,
general asymptotically AdS solutions should behave as
\bea
&& A(r)= a_1 r + a_0 + \cdots \,,\nn\\
&& h(r)= h(\infty) + h_3 e^{-3 A} + \cdots\,, \nn\\
&& F_{rt} (r)= 2q e^{-A}\,,
\label{bdinfty}
\eea
and the discussion of the boundary data for the scalar fields will be specified below.
The entropy, energy, charge densities and the temperature  have the expressions,
\bea
&&{\cal S}={2\pi\over \kappa^2}\, e^{2 A(r_H)}\,,\ \ \ {\cal E}=-{h_3\over \kappa^2 h(\infty)}\,, \nn\\
&& \rho ={2q\over \kappa^2 }\,, \ \ \ T={1\over 4\pi} e^{A(r_H)} {h'(r_H)\over \sqrt{h(\infty)}}\,.
\eea
In this computation all the thermodynamic quantities are measured with respect to the boundary time
$\sqrt{h(\infty)}\, t$ such that one may bring the boundary metric of (\ref{metric1}) in the standard form.

\section{Thermodynamic stability of the RN black brane}

In this section we shall discuss  the thermodynamic stability of the RN
black brane solution (\ref{ABJMcharged}). The thermodynamic stability is ensured if
the Hessian (second-derivative) matrix of the energy with respect to its thermodynamic variables
has no negative eigenvalues. With any negative eigenvalue, the system
becomes thermodynamically
unstable under small fluctuations that drive the system toward some
other stable point \cite{Gubser:2000mm,Hubeny:2002xn,Hirayama:2002hn}.
In the type IIB theory, $SO(6)$ R-charged black brane solution is available. But there it
is observed   that the R-charged black brane solution exhibits thermodynamic instabilities
at a temperature lower than a certain critical value \cite{Son:2006em}. The fate of black brane in this
unstable regime has not been known up to now.

Unlike the case of this type IIB counterpart, we find that our RN black brane solution
is thermodynamically stable. To show this, note first that the energy density can be
expressed as
\be
{\cal E}= {\kappa\, {\cal S}^{3\over 2}\over (2\pi)^{3\over 2}}  \Big(1+ {\pi^2\rho^2\over  {\cal S}^2} \Big)\,.
\ee
The components $H_{ij}$ of the Hessian matrix are given by
\bea
&&H_{11} =\frac{\partial^2{\cal E}}{\partial {\cal S}^2}\ =\  {3\kappa \over 4(2\pi)^{3\over 2}
{\cal S}^{1\over 2}} \Big(1+ {\pi^2\rho^2\over  {\cal S}^2} \Big)\,,\nn\\
&& H_{12}=\frac{\partial^2 {\cal E}}{\partial {\cal S}\partial  \rho}=-{\kappa \pi^2\over (2\pi)^{3\over 2}} {\rho\over {\cal S}^{3\over 2}}\,,\nn\\
&& H_{22} = \frac{\partial^2{\cal E}}{\partial \rho^2}\ =\ {2\kappa \pi^2\over (2\pi)^{3\over 2} {\cal S}^{1\over2}} \,.
\eea
The determinant of $H_{ij}$ then becomes
\be
{\rm det}\,\, H 
= {\kappa^2\over 16\pi {\cal S}} \Big(3+ {\pi^2\rho^2\over  {\cal S}^2} \Big)
\ee
together with ${\rm tr} \,\, H 
\ > \ 0$. This proves that the two eigenvalues are positive
definite. One may also consider turning on small quantities of the three further charges $\rho_a \,\,(a=1,2,3)$
in the Cartans of the $SU(4)$ R-symmetry. Including these contributions to the quadratic order, the energy density
has the expression,
\be
{\cal E}= {\kappa\, {\cal S}^{3\over 2}\over (2\pi)^{3\over 2}}  \Big(1+ {\pi^2 \over  {\cal S}^2} \, (\rho^2 +\rho_a\rho_a)\Big)\,.
\ee
Using now $5\times 5$ Hessian matrix, one may easily verify that
the system at $\rho_a=0$ is thermodynamically stable even in this enlarged space. Thus we conclude that the RN black brane
solution is thermodynamically stable.

\section{Geometrical instability of the RN black brane 
}
In this section, we shall investigate possible geometrical instabilities
of the RN black brane system with  
a particular probe 
mode turned on. Depending on the mass
of the supergravity mode, the RN black brane may become geometrically unstable
below a certain critical temperature $T_c$ leading to a new black brane solution wearing 
nontrivial hairs.

For scalar modes, arising of the instability may be understood as follows \cite{Horowitz einstein scalar,Horowitz criticality,Hartnoll:2008kx}.
Note that the usual geometrical stability
condition for the AdS$_{d+1}$ spacetime  is given by the so called Breitenlohner-Freedman (BF) bound \cite{Breitenlohner:1982bm},
\be
-{d^2\over 4} \
\le\  m^2\,,
\ee
which is indeed respected by any scalar mode of the present theory.
This is the relevant condition for the stability of the near boundary region of the RN black brane solution
that is asymptotic to AdS$_4$. On the other hand, the near horizon geometry of the extremal RN black brane
in $d+1$ dimensions is given by AdS$_2\times \mathbb{R}^{d-1}$ with a scaled AdS radius of $1/\sqrt{d(d-1)}$.
Hence for scalars, the BF bound of this region is violated if
\be
m^2 \ <\ -{d(d-1)\over 4} \,.
\ee
Therefore for our case of $d=3$, the instability below a certain temperature may occur
if the mass squared of a bulk scalar is ranged in
\be
-{9\over 4}\  \le\  m^2 \ < \ -{3\over 2}\,.
\ee
For the higher spin fields with spin $s\ge {1\over 2}$, one may show that there is no
potential instability once they are neutral under the $U(1)$ gauge field of the RN
black brane. Thus scanning the supergravity modes in Table \ref{tableone},
one finds that the possible gravitational instabilities are limited to the case of scalar fields
with $m^2=-2$.  As explained in Section 3, these bulk scalars are dual to the field theory
operators of dimension $\Delta =1$ and $2$.

To show the instability of the black hole,
let us study the positivity condition of the energy functional for the scalar field.
The energy of a scalar field reads
\be
E=
\int dr dx dy \sqrt{-g} \bigg[ |g^{tt}| (\dot{\phi})^2 + g^{rr} (\phi')^2 + g^{xx} (\partial_x\phi)^2 + g^{yy} (\partial_y\phi)^2 + m^2\phi^2 \bigg]\,,
\ee
where 
dots and 
primes
denote derivatives with respect to $t$ and $r$ respectively.
If there exists any normalizable (probe) mode $\varphi$ of
the scalar field $\phi$ which makes this energy functional negative,
the geometrical instability of the RN background can be triggered driving the system to some new stable configuration.
In order to find a possible mode of instability,
we take $\varphi$ as a function of $r$ only and look for the negative fluctuation mode
satisfying 
\begin{eqnarray}
\left( e^{3A}h(r) \varphi' \right)' + 2 e^{3 A} \varphi - \kappa_0^2 \, \varphi = 0,
\label{stabilty eom}
\end{eqnarray}
where $\kappa_0$ is a real constant and we set $m^2=-2$.
Of course one may turn on the spatial fluctuation by considering the probe field depending on $x$ and $y$ by
 $e^{i(k_x x+ k_y y)}$ but this will only increase the energy of the system. Hence the above consideration
 will be sufficient.

The boundary conditions are  crucial for a determination of the
solution of  (\ref{stabilty eom}).
For $r=\infty$, we note that
the behavior of scalar fields in the near boundary region
takes a form,
\begin{eqnarray}
\varphi \sim s_{\Delta}(x) e^{-(3-\Delta)A(r)} + o_{\Delta}(x) e^{-\Delta A(r)}+ \cdots,
\end{eqnarray}
where $\cdots$ denote higher order terms.
From the standard dictionary of AdS/CFT, the presence of  $s_\Delta$ corresponds to turning on an external source term
for the dual operator $O_{\Delta}(x)$ while $o_\Delta$ represents  the operator expectation value
$\langle O_{\Delta}(x)\rangle_{s_\Delta}$
in the presence of the source $s_\Delta(x)$. In our present problem, we would like to consider the system without
introducing the source term and, thus, our choice of the boundary condition here and below is
$s_\Delta=0$ at
$r=\infty$.
The other boundary condition is for the horizon of black brane.
Basically we shall require nonsingularity of the configuration there. For this,
we need to expand the equation (\ref{stabilty eom}) into a generalized power series in $r-r_H$,
and
then we can obtain boundary condition for the horizon as 
\begin{eqnarray}
\varphi'(r_H)&=& -\frac{2 -\kappa_0^2 e^{-3 A(r_H)}}{h'(r_H)}\varphi(r_H)\,,\nonumber\\
\varphi''(r_H)&=& \frac{ 2-\kappa_0^2 e^{-3 A(r_H)}}{2 \left(h'(r_H)\right)^2}
\left( 3 A'(r_H)h'(r_H)+ h''(r_H)+2 -\kappa_0^2 e^{-3A(r_H)} \right)\varphi(r_H)\nn\\
&-&\frac{ 3\kappa_0^2\, A'(r_H)e^{-3A(r_H)}}{2 h'(r_H)}\varphi(r_H),
\end{eqnarray}
by setting the coefficients of the negative powers of $r-r_H$ to zero.
Note that the denominators are given by $h'(r_H)$: If it is too small, we may have trouble
in numerical computations. In order to avoid potential numerical instabilities,
we use another form of RN solution given by
\begin{eqnarray}
&&A(r) = ( 3 - q^2 )r,~~h(r)=\frac{ 1-(1+q^2) e^{-3A(r)} + q^2 e^{-4 A(r)} }{(3-q^2)^2},\nonumber\\&&A_t = -
\frac{2q}{3-q^2}\left(e^{-A(r)} -1\right)\,, \ \ \ \phi_a=0\,.
\label{rnsol}
\end{eqnarray}
This solution can be related to (\ref{ABJMcharged}) by the coordinate transformation,
\bea
&& r' = {r-r_H\over 3 - q^2e^{-4 r_H}} ,~~t'=e^{r_H}(3 - q^2e^{-4 r_H})\, t\nonumber\\&& x' = e^{r_H}\,x
\,, \ \ \ \ \ \ y' = e^{r_H}\,y\,,
\label{ct}
\eea
with the redefinition of the parameters,
\be
q'=q\, e^{-2r_H}\,, \ \ \ \ \epsilon'=\epsilon\, e^{-3r_H}\,.
\label{pt}
\ee
After the transformation, we drop primes for the notational simplicity.
The temperature for this system reads
\be
T={1\over 4\pi} {({3-q^2})}\,,
\ee
and the location of the horizon is at $r=0$. Here and below, we shall use this rescaled background for the
numerical analysis.

Now one can solve the probe equation (\ref{stabilty eom})  numerically to find the
temperature below which the negative mode begins to develop. For the numerical analysis,
we use the standard shooting method based on a Mathematica coding.
The results are as follows: 
The on-set of instability  occurs at
\be
T_c=0.0395(7)  \ \ \ \big[\,q_c=1.582(0)\,\big]
\label{ins1}
\ee
for the $\Delta=1$ operator.
For the $\Delta=2$ operators, we found   
\be
T_c=0.0003(5)  \ \ \ \big[\,q_c=1.7307(7)\,\big]\,.
\label{ins2}
\ee
The differences in the numbers of significant digits of $T_c$ and $q_c$ arise
due to the fact that $T_c$ is proportional to $3-q_c^2$.


\section{Phase structures and critical exponents}

In the previous section, we have established the instability of the RN black brane in the
presence of bulk scalars with $m^2=-2$. As temperature decreases below the critical temperature,
the scalar fields begin to develop a nontrivial profile that affects the original RN black brane
geometry.
As we shall see later on, the boundary CFT 
undergoes a phase transition
by the condensation of the expectation value of the dual field theory
operator $O_{\Delta}(x)$.

In this section,  we shall investigate this phase transition by looking at the
supergravity solutions.
Since the changes in the thermodynamic quantities like the
entropy, energy and so on are encoded in the geometry, the probe analysis of the previous section
alone is not sufficient 
and  inclusion of the full back-reaction to the geometry will be essential. 
Our study of the relevant solutions will be mainly based on numerical analysis.

Our starting point of the analysis\footnote{We use here our original coordinates without performing the coordinate
transformation in (\ref{ct}) and (\ref{pt}).
Then, when the scalar field vanishes, we shall directly
obtain
the exact RN  solution (\ref{rnsol}) by
fixing some freedoms in our coordinate choice.}
is the set of equations in
(\ref{fund}) and (\ref{eqom}) where we turn on only one scalar field $\phi$
with $m^2=-2$.
To set up the problem completely, one has to also specify
the boundary conditions. Let us begin with the horizon side. At the horizon, $h(r)$ has to be zero
at some finite $r=r_H$ which is the coordinate location of the horizon.  Using the translational
freedom of the $r$ coordinate, we shall choose
\be
r_H=0\,.
\label{bd0}
\ee
In addition, we note that one has the scaling freedom for
$x_\mu=(t,x,y)$ and the coordinate $r$ 
to generate a new solution. Using this
we shall fix
\begin{eqnarray}
A(0)=0~~~{\rm and}~~~h'(0)=1.
\label{bd3}
\end{eqnarray}
At the horizon we basically require that
all the fields in (\ref{eqom}) should be regular. Setting $\phi(0)=u$, the regularity leads to
\begin{eqnarray}
\phi'(0)&=& - 
\frac{2\, u}{h'(0)}~,\nn\\
A'(0)&=&\frac{3- q^2 e^{-4 A(0)}+ u^2}{h'(0)}\,\,,
\label{bd1}
\end{eqnarray}
and
\begin{eqnarray}
h''(0) &=& -9 -3 u^2 + 7 q^2 e^{-4 A(0)}~,\nn\\
A''(0) &=& - \frac{2\,u^2}{\left(h'(0)\right)^2}\,\,,\nn\\
\phi''(0) &=&  
\frac{2\, u}{\left(h'(0)\right)^2}\left( 1 + 2 q^2 e^{-4 A(0)}\right)\,.
\label{bd2}
\end{eqnarray}
($\,$Solving (\ref{fund}) and (\ref{eqom}) with the coordinate and boundary
conditions
in (\ref{bd0})-(\ref{bd2}) for the case of vanishing scalar field
leads directly to the exact RN solution in (\ref{rnsol}) without
need of the transformation in (\ref{ct}) and (\ref{pt}).)
At $r=\infty$,  we shall impose the behaviors of fields in (\ref{bdinfty}) that are required for the
asymptotically AdS spacetime.
Again note that the scalar field in the near boundary region takes the form
\begin{eqnarray}
\phi \sim s_{\Delta}(x) e^{-(3-\Delta)A(r)} + o_{\Delta}(x) e^{-\Delta A(r)}+ \cdots,
\end{eqnarray}
where $\cdots$ denote higher order terms. (The interpretation of $s_{\Delta}(x)$ and $o_{\Delta}(x)$
in the boundary CFT is the same as that in the previous section.)
We shall set the source term $s_{\Delta}(x)=0$,
which corresponds to the boundary system without an external source term. 
By this last
condition, $u$ will be determined as a function of $q$ by $u=u(q)$.

\begin{figure}[ht]
\center
\includegraphics[width=7.2cm]{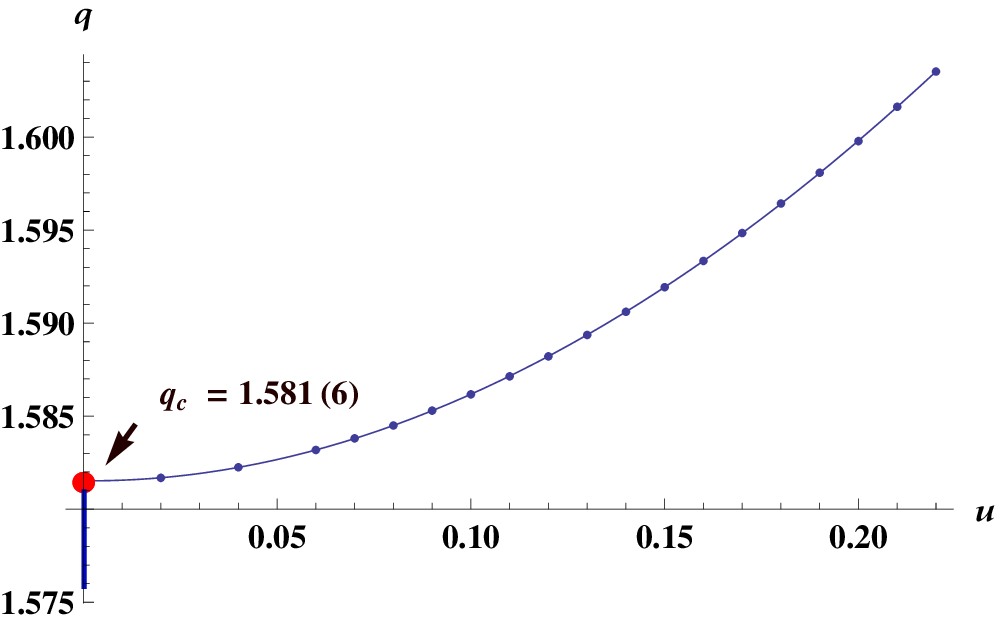}
\includegraphics[width=7.5cm]{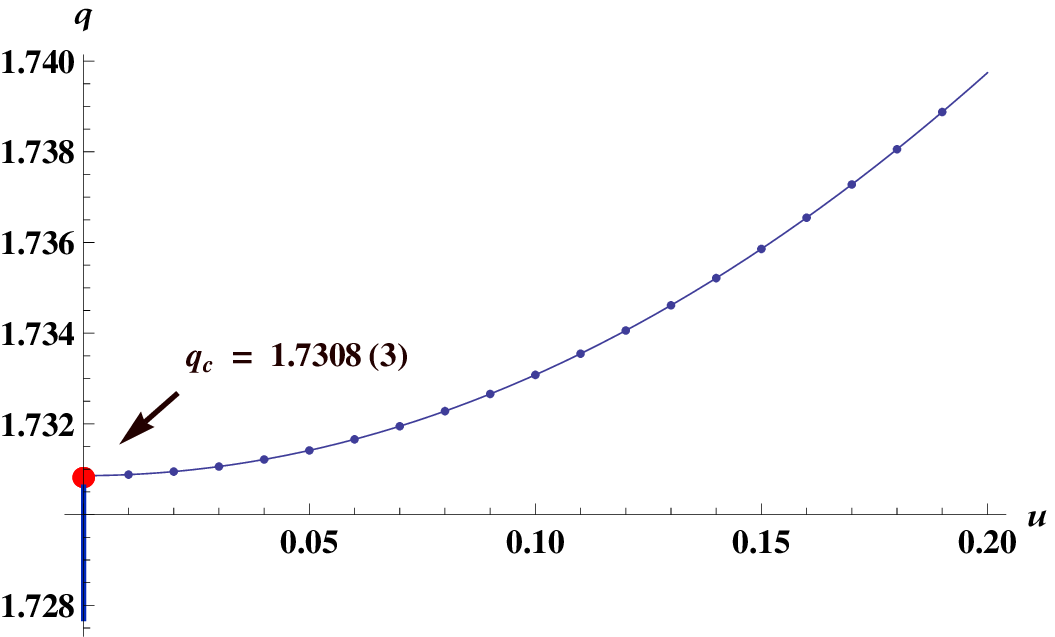}
\caption{\small
The functions $u(q)$ in $(u,\,\, q)$ plane are depicted
in the left and the right sides respectively for  the $\Delta=1$ and the $\Delta=2$ cases.
For each case, the development of the scalar profile is represented by nonvanishing $u(q)$
for $q > q_c$; The dots represent our numerical data set while the solid curve is for the
fitting function
obtained by the standard curve fitting method in Mathematica. Below $q_c$, $u(q)=0$ is
indicated by the blue solid line,
which corresponds to the RN black holes in (\ref{rnsol}).
In addition, we marked the numerical values of  $q_c$ by the red circles, which are
 given by 1.581(6)  and 1.7308(3) 
 respectively for the $\Delta =1$ and the 
 $\Delta =2$ cases.
}
\label{f4}~~
\end{figure}

Now adopting the shooting method based on a Mathematica coding,
we perform a numerical analysis for $\Delta=1$ and
$2$ cases separately. For each case, we find one set of solutions that is parameterized
by the value of $q$. 
The resulting functions 
$u(q)$ in $(u\,,\,q)$ plane
are drawn in Fig.~\ref{f4}.
For each case, there is a critical value of $q_c$ beyond which  the corresponding scalar field
begins to develop a nontrivial profile. Namely if $q\ \le\  q_c$, the exact RN solution
in (\ref{rnsol}), which satisfies all the coordinate and
boundary conditions in (\ref{bd0})-(\ref{bd2}),
remains intact. This part is depicted in Fig.1 by the vertical blue solid line
for $q \le q_c$.
If $q \ >\ q_c$, the RN black brane
is modified by wearing a nontrivial scalar hair with $u(q)\neq 0$; In Fig.1, the blue dots
represent the data set which we obtained by the numerical analysis. The blue solid curve
represents the fitting function that is obtained by the standard curve fitting method in
Mathematica.

Thus the critical temperature may be evaluated
by using the RN black brane solution (\ref{rnsol}) with $q=q_c$ leading to
\be
T_c= {1\over 4\pi}\, ({3-q_c^2})\,.
\ee
By the standard curve fitting method, we found
\be
T_c=0.0396(4)  \ \ \ \big[\,q_c=1.581(6)\,\big]\,,
\label{tceq1}
\ee
for the $\Delta=1$ scalar field,
and
\be
T_c=0.0003(4)  \ \ \ \big[\,q_c=1.7308(3)\,\big]\,
\label{tceq2}
\ee
for the $\Delta=2$ scalar fields. In principle, these critical values  should agree precisely with
those from the probe analysis of the previous section.
Therefore the numerical precision of our analysis can be estimated by comparison of
the numerical values from the two methods. 
  We see that the $\Delta =2$ critical temperature,
whose value is closer to
zero,  has a poorer numerical accuracy.

In order to show further numerics, instead of using the charge density $\rho=2q/\kappa^2$,
we shall use
simply  $q$ which differs from the
actual charge density by a constant factor $2/\kappa^2$.
($\,$See (\ref{newton}) for the definition of
$\kappa$.)
In the above sets of solutions, both $q$ (or the charge density) and the
temperature $T=T(q)$ changes as one changes $q$ along the transition.
But what we want  is to fix $q$ (or the charge density) while changing the
temperature of the system along the transition. In order to generate such  sets of
solutions, we use the coordinate transformation of the gravity system by
\be
x^\mu \ \rightarrow \ {x^\mu / \lambda_s}\,.
\ee
Note that, by the scale transformation,  the thermodynamic quantities transform as
\bea
&& T\ \rightarrow \ \lambda_s \, T\,, \ \ \  q\ \rightarrow \  \lambda_s^2 \, q\,, \ \ \ {\cal E}\ \rightarrow \
\lambda_s^3{\cal E}\,,\nn\\
&& {\cal S}\ \rightarrow \ \lambda_s^2 {\cal S}\,, \ \ \ o_\Delta \ \rightarrow\ \lambda_s^{\Delta} o_\Delta
\,, \ \ \ s_\Delta \ \rightarrow\ \lambda_s^{3-\Delta} s_\Delta\,,
\label{scalet}
\eea
and, hence,
\be
{\partial o_\Delta\over \partial s_{\Delta}} \ \rightarrow\ \lambda_s^{2\Delta -3} \,\,
{\partial o_\Delta\over \partial s_{\Delta}}\,.
\ee
Using this scaling transformation, we shall fix $\tilde{q}=1$ by choosing $\lambda_s=1/\sqrt{q}$.
(\,The quantities carrying a tilde are the ones after the scale transformation (\ref{scalet}).\,)
The new sets
of scaled solutions are now parameterized by the rescaled temperature
\be
\tilde{T}= {T\over \sqrt{q}}
\ee
with fixed charge density $\tilde{\rho}=2/\kappa^2$. Using the values in (\ref{tceq1}),
we found the rescaled critical temperatures $\tilde{T}_c=T_c/\sqrt{q_c}$ as 
\be
\tilde{T}_c=0.0315(6)
\label{ttc1}
\ee
for the $\Delta=1$ scalar field.  For the $\Delta=2$ scalar fields, we found
\be
\tilde{T}_c=0.0002(5)
\label{ttc2}
\ee
using the values in (\ref{tceq2}).

\begin{figure}[ht]
\center
\includegraphics[width=7.5cm]{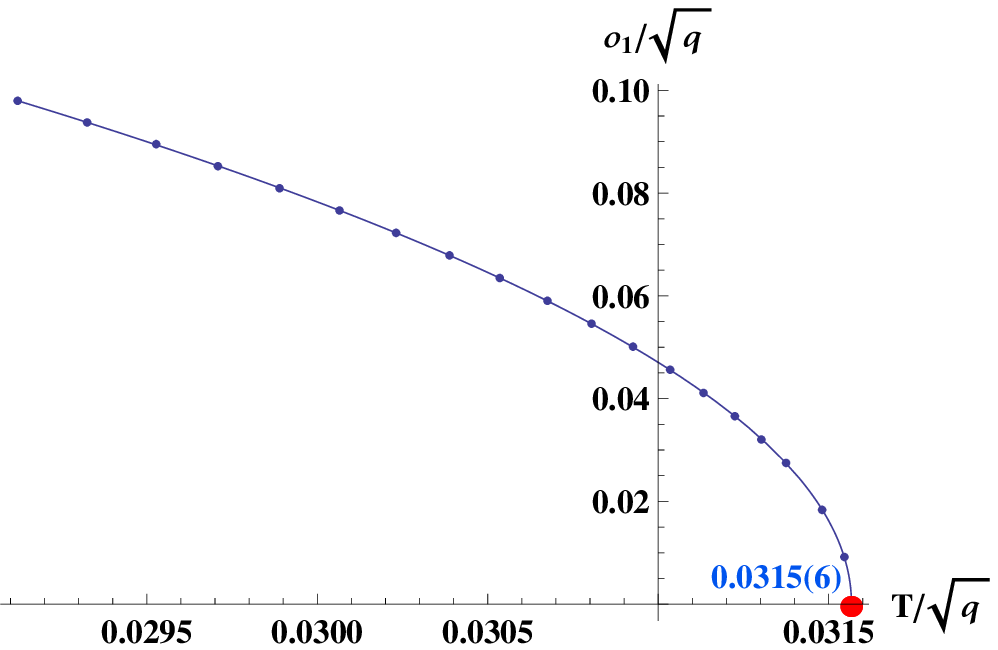}
\includegraphics[width=7.5cm]{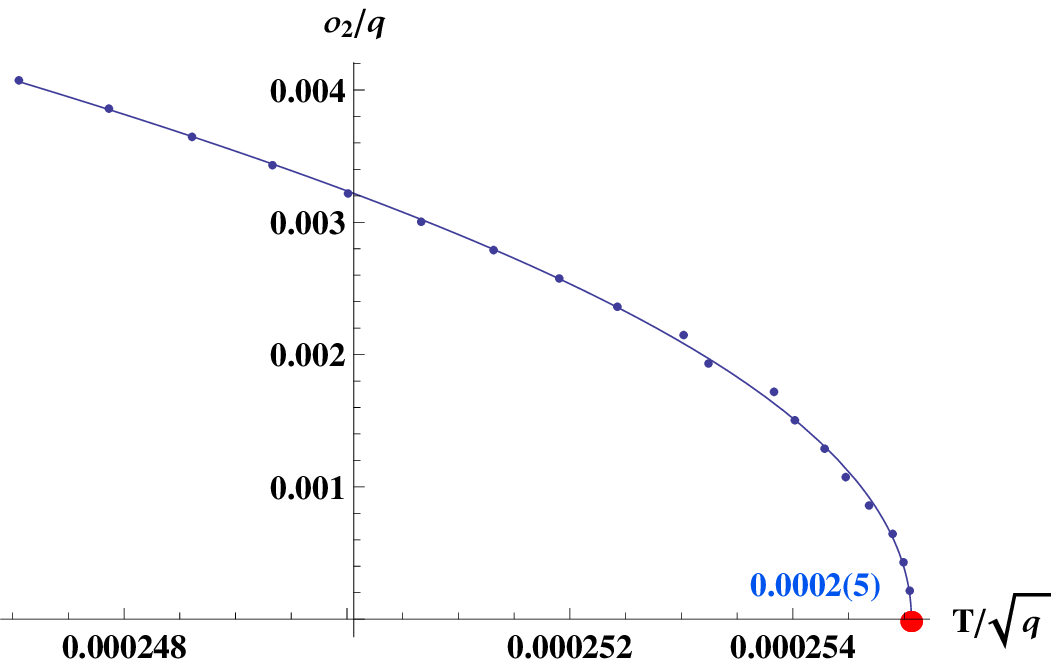}
\caption{\small
The phase diagrams in the left and the right sides are
respectively for the $\Delta=1$ and the $\Delta=2$ scalars.  The system undergoes a symmetry
breaking transition from a symmetric phase to  a broken phase as
the temperature is lowered below the critical temperature. For each case, the blue dots represent
our numerical data set and the blue solid curve depicts the fitting function that is obtained by the
standard curve fitting method in Mathematica.
We marked the numerical values of  $\tilde{T}_c$ by the red circles, which are
 given by 0.315(6)  and 0.0002(5) [see (\ref{ttc1}) and (\ref{ttc2})]
 respectively for the $\Delta =1$ and the 
 $\Delta =2$ cases.
}
\label{f5}
\end{figure}

Below the critical temperature, one finds a development of expectation value $o_\Delta$ which signals
a phase transition. As we shall see details later on, this basically corresponds to the spontaneous symmetry breaking
transition of the $SU(4)$ R-symmetry in which the condensate $o_\Delta$ plays the role of the order parameter.
In terms of the scaled variable $\tilde{o}_\Delta=o_\Delta/q^{\Delta/2}$, the phase diagrams are depicted
in Fig.~\ref{f5} for the $\Delta=1$ and the $\Delta=2$ cases.

As expected, the transition for $\Delta=1$ occurs at a higher critical temperature. Then the transition for
$\Delta=2$ cannot be treated separately. Namely one has to turn on both of the $\Delta=1,\,\,2$
scalar fields 
around the region of the $\Delta=2$ transition in which the $\Delta=1$ scalar field has developed
some finite amount of profile representing the condensation. However remember that
the scalar part of our starting Lagrangian is only valid up to quadratic orders. Hence our treatment
 loses its validity around the region of the $\Delta=2$ transition. Though it is an interesting problem
 to clarify further, we shall leave it to the future investigation. For the remaining we shall be  focusing
 on the transition involving the $\Delta=1$ condensation.

 The phase transition is of second order as in the usual cases of symmetry breaking
 transition. We note that natures of phase transitions are in general
 classified by their critical exponents.  Here we  compute numerically the exponents
 $\alpha$, $\beta$ and $\gamma$ respectively defined by
 \bea
&&{\partial {\cal E}\over \partial T}\ \sim \  |T-T_c|^{-\alpha} \nn\\
&& o_1\ \sim \  |T-T_c|^{\beta} \nn\\
&& \chi_1 = {\partial {o_1}\over \partial s_1}\Big|_{s_1=0}\ \sim  \  |T-T_c|^{-\gamma}\,.
\label{chi1}
\eea

\begin{figure}[ht]
\center
\includegraphics[width=7.5cm]{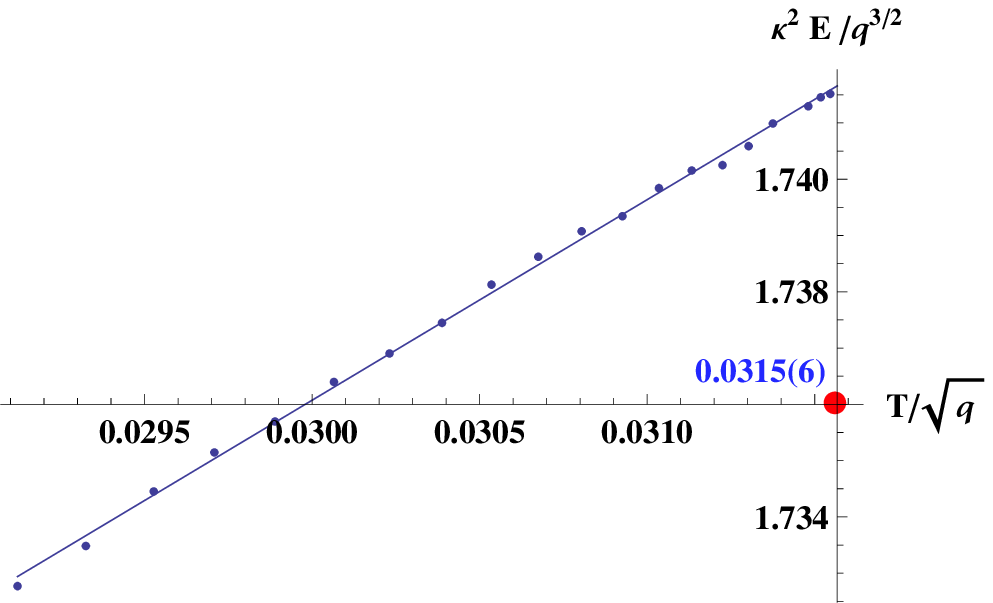}
\includegraphics[width=7.5cm]{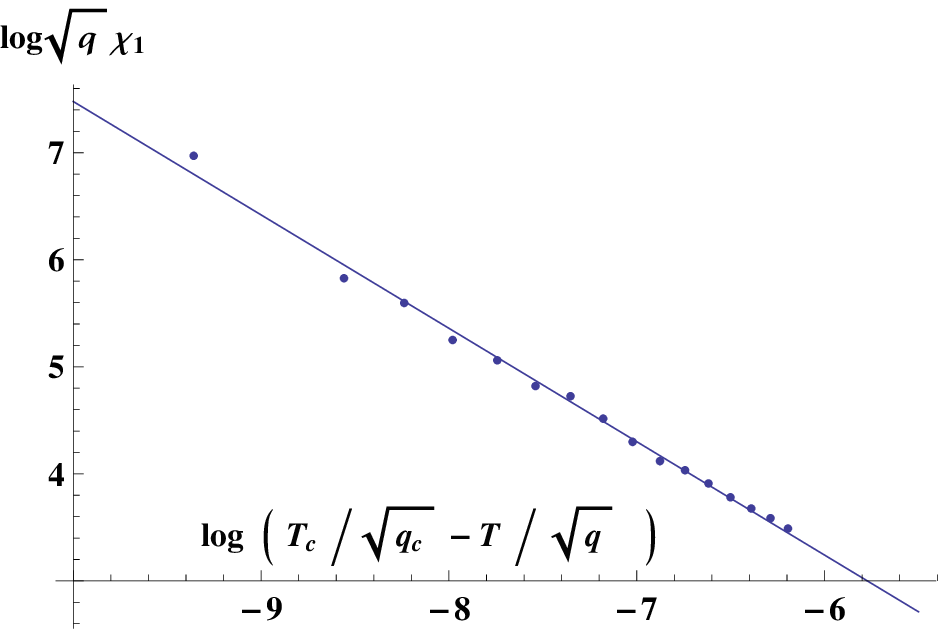}
\caption{\small
The left figure shows the scaled energy density as a function of temperature.
The right hand side shows $\log \Big(\sqrt{q}\chi_1\Big)$  in (\ref{chi1}) with respect to
$\log \Big( T/\sqrt{q}-T_c/\sqrt{q_c}
\Big)$. For each case, the blue dots represent
our numerical data set and the blue solid curve depicts the fitting function that is obtained by the
standard curve fitting method in Mathematica.
}\label{f6}
\end{figure}

For the numerical estimation of the exponents $\alpha$, $\beta$ and $\gamma$, we use the numerical data sets
respectively
of the forms ($\log (T-T_c)$, $\log (\partial E/\partial T)$), ($\log (T-T_c)$, $\log o_1$) and ($\log (T-T_c)$,
$\log \chi_1$). Using the fitting with linear least squares, we estimated the relevant slopes as well as
their standard deviations.
From the corresponding data set for 
 $o_1$ in Fig.~\ref{f5}\,, one finds
 \be
 \beta_n=0.04978\pm 0.0032\,,
 \ee
 which is in a good  agreement with the mean field value $\beta=1/2$.
 Here and below the subscript $n$ indicates that the relevant exponent is obtained by the
 numerical analysis.
 The left side
of Fig.~\ref{f6} shows the behavior of the energy as a function of temperature
in the vicinity of the transition region. Form the corresponding data set, the exponent for the
specific heat  is identified as
\be
\alpha_n=0.012\pm 0.018\,,
\ee
which agrees well with the mean field value $\alpha=0$.
The right hand side of
Fig.~\ref{f6} shows the temperature dependence of the susceptibility $\log \chi_1$
with respect to the logarithm of temperature.
Again from  the corresponding numerical data set, one finds 
\be
\gamma_n=1.011\pm 0.015\,,
\ee
which agrees well with the mean field value $\gamma=1$.
One can check  that
\be
(\alpha+2\beta+\gamma)_n=2.019\pm 0.024
\ee
which is consistent
with
the so called Rushbrooke scaling law
\be
\alpha+ 2\beta +\gamma=2\,.
\ee



Finally we shall reconfirm the above exponents
by the analytic treatment. We note first that,
due to the boundary condition $\phi(0)=u$, the scalar field behaves as $\phi \propto u$
when $u\ \ll \  1$.  Then inspecting the equations of motion
(\ref{eqom}) together with the boundary conditions (\ref{bd1}) and (\ref{bd2}), one finds that
$s_1$ and $o_1$ should behave as
\bea
 s_1 &=& u \,\, \big(\, a_0 (q) +a_2(q) u^2 + O(u^3)\, \big)
\nn\\
 o_1 &=& u\,\, \big(\, b_0(q)+b_2(q) u^2 +O(u^3)\,\big)\,,
\label{so}
\eea
before imposing the last boundary condition of $s_1=0$.
For $|u|\ll 1$, the coefficients of the $u^3$ terms in $s_1$ and $o_1$
are basically controlled by the $u^2$ terms of $A'(0)$ and $h'(0)$ in ({\ref{bd1}})
 and ({\ref{bd3}}), which one may argue by
 a careful examination of the equations of motion in (\ref{eqom}): By this consideration, one may show that
\bea
a_2(q)= a_{20} +O(q-q_c)\,,\ \ \ \ b_2(q)= b_{20} +O(q-q_c)
\eea
with $a_{20}\neq 0$ and $b_{20}\neq 0$.
 Then, since $s_1=0$ has a nontrivial solution $u(q)\neq 0$ only for $q > q_c$,
$a_0(q)$  should behave as
\bea
a_0(q)= a_{01}\, (q-q_c) + O[\,(q-q_c)^2]
\eea
with $a_{01} a_{20} < 0$
for $|q-q_c| \ll 1$. (Here we assume the existence of $q_c$, which is justified by
our numerical analysis.)
Since we know that the $o_1=0$  boundary condition instead of $s_1=0$ leads to a different value for
the critical charge, $b_0(q)$ has the expansion of the form
\be
b_0(q)= b_{00} +O(q-q_c)
\ee
with $b_{00}\neq 0$.
This argument shows that the $s_1=0$ condition leads to a 
solution
\bea
u=
\left[
\begin{array}{cr}
 \sqrt{-a_{01}\over a_{20}}\,\, (q-q_c)^{1\over 2} \, (1+O(q-q_c)) 
 &    \ \ \ \  {q\ \ge\ q_c}\phantom{a}\\
 0 &  {q \ < \ q_c}\,\,.
\end{array}
\right.
\eea
Thus,
\be
o_1\ \sim \ |T-T_c|^{1\over 2}\,,
\ee
which implies that $\beta={1\over 2}$. Since the scalar field contribution to the energy density is of order
$u^2$,  
the energy density has to be of the form
\be
{\cal E} = e_0 + e_1 (q-q_c) + e_2 u^2 + \cdots
\ee
where the first two terms are from the original RN black brane with $u=0$.
We then conclude that $\alpha=0$ since the leading power of specific heat is zero.
By changing $u=u(q)$ to $u=u(q)+\delta u$, $s_1$ and $o_1$ in (\ref{so}) vary by
\be
\delta s_1 =3 a_{20}(u(q))^2 \delta u\,, \ \ \ \
\delta o_1= b_{00}\delta u
\ee
to the leading orders.
Then the susceptibility behaves as
\be
{\delta o_1\over \delta s_1} = {b_{00}\over 3 a_{20}} {1\over (u(q))^2} \ \sim \ |T-T_c|^{-1}\,,
\ee
which implies that $\gamma=1$. This proves that our exponents are those of the mean field theory.

\section{Interpretations and discussions}
In the previous section, we have established the phase
transition whose exponents
belong to the universality class of the mean field
theory. 
For the resulting expectation value of $\Delta=1$ operator, let us introduce
a notation
\be
{\cal M}_{IJ}(x) = \langle O^I_J (x)\rangle
\ee
where $O^I_J$ is given in (\ref{op1}).
Since ${\cal M}_{IJ}= {\cal M}^*_{JI}$ and ${\cal M}_{I\,I}=0$, the $4\times 4$ matrix ${\cal M}$
is Hermitian and traceless.
We have seen that the boundary CFT undergoes a phase transition. Above the critical temperature $T_c$,
the condensate is vanishing, i.e.  ${\cal M}=0$ while at lower temperature, ${\cal M}\neq0$.
 This is the spontaneous symmetry breaking phase transition where
the $SU(4)$ R-symmetry of the ABJM theory is broken by the presence of the
condensate ${\cal M}$.

Using ${\cal M}$ as an order parameter, the phase transition may be effectively described
by the Landau free energy $F$
\be
F/T=\int d^2\vec{x} \Big(
\,\,\, {\tr}\, \nabla {\cal M} \cdot \nabla {\cal M} -{1\over T}{\tr }\, {\cal M} {\cal H}
+2 r_0 (T-T_c) {\tr }\, {\cal M}^2 + g_0 ({\tr } {\cal M}^2)^2
\Big)\,,
\ee
where $\vec{x}=(x,y)$ and the $4\times 4$  traceless Hermitian matrix ${\cal H}$
is for the external source term in the adjoint representation of $SU(4)$.
The form of the quartic term is determined by the symmetry. Due to the symmetry of the gravity
solution, it should be invariant under the transformation of    ${\cal M} $ by the
SU(4) R-symmetry.
The Landau free energy is minimized
if
\bea
{\cal M}= \left[
\begin{array}{lr}
 0 &  {T \ > \ T_c}\\
 \sqrt{r_0\over g_0}\, (T_c-T)^{1\over 2}\,\hat{n} &  \ \ \ \ \ \ \ \  \ \ {T\ \le\ T_c}
\end{array}
\right.
\eea
where $\hat{n}$ is a $4\times 4$ traceless Hermitian matrix with $\tr\, {\hat{n}}^2=1$.
This implies that $\beta=1$. Noting $C_{V}=-T{\partial^2 F\over\partial T^2}$ with
\be
F \sim (T-T_c)^2\,,
\ee
one has $\alpha=0$. Similarly one can  check that $\gamma =1$ from the definition of the
susceptibility.  One can further compute the correlation length scale
$\xi$, 
\be
\xi = (\, 2 r_0(T_c-T)\,)^{-{1\over 2}}\ \sim \ |T-T_c|^{-\nu}\ \ \ \ \ \ (\,{\rm for} \ \ {T\ \le \ T_c}\,)\,\,,
\ee
from which one finds $\nu=1/2$.
It will be interesting to test the prediction of the mean field theory
by the direct analysis of the bulk gravity system. But on top of the new background obtained
from the numerical analysis, further numerical analysis of solving the scalar fluctuation
equation has to be performed. However this requires some improvement of  the current method
to reach the required numerical precision.

As we described in Section 3, our $U(1)$ charged black brane describes a system  where
the $U(1)$ number is carried by both bosonic and fermionic degrees.
 The build-up of the fermi surface at weak coupling side
 would not be possible if the bosonic degrees were
 in a unbroken phase.  Indeed we have argued that
 there will be a condensation of the elementary degrees at weak coupling with low enough
 temperature.
 For the strong coupling side, we know that  the  basic
 degrees that are weakly coupled among themselves are now organized by
 the supergravity modes.
 We found that the symmetric phase of the RN black brane
 again becomes unstable when $T\ < \ T_c$. The effective mass squared of the Landau free energy
 description becomes negative in this region driving the
 symmetry breaking phase transition by the condensation of the operator expectation value.
 This partially explains the fate of the bosonic $U(1)$ charged state of the ABJM theory
 at strong coupling. But as observed before, there is a finite
 entropy of the extremal RN black brane  at zero temperature. However due to the
 phase transition, ordering of degrees will lead  to a reduction of entropy
 in general. Hence there can be a chance of having a
 zero-entropy system at zero temperature.  One interesting point of study is then
 the modification of the RN black brane by the scalar condensation
in the near zero temperature region. Since the scalar part of our starting Lagrangian
is only valid up to quadratic orders, we cannot answer this question
at the moment.
One has to improve our understanding
of higher-order scalar contribution to the Lagrangian first.

Another interesting aspect is the build-up of the fermi surface
as mentioned before. The picture presented in Ref.~\cite{BakRey2} is seemingly
plausible, which mostly concerns the region of near
zero temperature, and works only for a pure RN black brane system.
But now close to zero temperature, our black brane solution
is significantly modified by the condensation of nontrivial scalar profile.
Further studies of the low temperature region appear interesting, in particular with focus
on the question whether the Fermi picture is still valid.

Finally we comment on the theorem by Coleman-Mermin-Wagner-Hohenberg \cite{Coleman:1973ci}. The theorem
states that
a continuous global symmetry cannot be broken spontaneously
in 1+1 dimensions at zero temperature and 2+1 dimensions  at finite temperature.
Our example appears to be contradicting with what the above theorem states.
However, we are working in the strict large $N$ limit, which makes the dual gravity
system completely classical. This violates some assumptions of the above theorem
as pointed out in Ref.~\cite{Anninos:2010sq}. Our example here is reminiscent of
the clash of the unitarity in the explicit time dependent black hole
solution which describes a thermalization of a boundary field theory \cite{Bak:2007qw}.
There again the trouble
stems from the large $N$ limit of the boundary field theory.


\section*{Acknowledgement}
K.K. Kim would like to thank Gungwon Kang for helpful discussions. DB
was supported in part by
 NRF SRC-CQUeST-2005-0049409 and  NRF Mid-career Researcher Program 2011-0013228.
KK and SY were supported in part by WCU Grant No. R32-2008-000-10130-0.



\end{document}